\documentclass[a4paper,11pt]{article}
\usepackage{nfraconf,graphicx,cite,amssymb}

\usepackage{times}

\begin{document}
\baselineskip=13pt

\title{GIANT AND `DOUBLE-DOUBLE' RADIO GALAXIES: IMPLICATIONS FOR THE EVOLUTION OF POWERFUL RADIO SOURCES AND THE IGM}

\author{A.P. Schoenmakers\footnote{Current address: N.F.R.A., P.O. Box 2, 7990-AA, Dwingeloo, The Netherlands}}
\address{Astronomical Institute Utrecht, P.O. Box 80000, 3508-TA Utrecht,\\
and Sterrewacht Leiden, P.O. Box 9500, 2300-RA Leiden,\\ 
The Netherlands; E-mail: schoenma@astro.uu.nl}

\author{A.G. de Bruyn}
\address{N.F.R.A., P.O. Box 2, 7990-AA, Dwingeloo,\\
and Kapteyn Astronomical Institute, P.O. Box 900, 9700-AV, Groningen,\\
The Netherlands; E-mail: ger@nfra.nl}

\author{H.J.A. R\"ottgering}
\address{Sterrewacht Leiden, P.O. Box 9500, 2300-RA Leiden,\\ 
The Netherlands; E-mail: rottgeri@strw.leidenuniv.nl}

\author{H. van der Laan}
\address{Astronomical Institute Utrecht, P.O. Box 80000, 3508-TA Utrecht,\\
The Netherlands; E-mail: vdlaan@astro.uu.nl}
\author{K.-H. Mack}
\address{Istituto di Radioastronomia del CNR, Via P. Gobetti 101, I-40129 Bologna, Italy\\ E-mail: mack@astbo1.bo.cnr.it}
\author{C.R. Kaiser}
\address{Max-Planck-Institut f\"{u}r Astrophysik, Karl-Schwarzschild-Str. 1,
85740 Garching bei M\"{u}nchen,\\Germany; E-mail: kaiser@mpa-garching.mpg.de}
\maketitle

\abstract{Giant radio sources form the linear size extreme of the
extra-galactic radio source population. Using the WENSS survey, we
have selected a complete sample of these sources.  We have
investigated the properties of their radio structures. 
We find, among other things, that these
sources are old (50-100 Myr) and have higher advance velocities than
smaller sources of similar radio power. We find pressure gradients in
their radio lobes, suggesting that the lobes are still overpressured
with respect to the environment. Further, we find no evidence for a
cosmological evolution of the radio lobe pressures with increasing
redshift, at least up to $z\sim 0.4$, other than that caused by
selection effects.  We argue that a much fainter sample of giant
sources than currently available is needed to constrain the pressure
in their environments, the IGM, and that SKA can play an important
role in studying such sources. Another extremely important discovery
is that of a population of radio sources with a so-called
`double-double' structure, i.e. that of a small two-sided radio source
embedded inside a much larger two-sided structure. We argue that these
sources result from an interrupted central jet-forming activity. As
such, they are the most convincing examples of radio sources with a
history of interrupted activity, yet. Since the inner lobes advance
within the outer lobes, high resolution low frequency ($\sim 200$ MHz)
polarization studies may reveal the constituents of radio lobes and
cocoons. We thus argue for a SKA design that can provide low-frequency
images at arcsec resolution, but which is also sensitive to structures
as large as a few tens of arcminute on the sky.}

\section{Giant Radio Galaxies}

The central activity in radio loud Active
Galactic Nuclei (AGN) produces relativistic outflows of
matter, the so-called `jets', for a prolonged period of time, possibly
up to a few $10^8$ yr.  These jets, when powerful enough, inflate a
cocoon (e.g. \cite{scheuer74}, \cite{falle91}) which expands first in the
Interstellar Medium (ISM) and later in the Intergalactic Medium
(IGM). Within this cocoon, which exists of accelerated jet material,
synchrotron radio emission is produced.  The evolution of the cocoon
can therefore be traced by observations of the radio lobes.

Giant radio galaxies (GRGs) are radio sources whose lobes span a
(projected) distance of above 1 Mpc\footnote{We use $H_0 =
50$~km\,s$^{-1}$\,Mpc$^{-1}$ and $q_0 = 0.5$ throughout this
contribution.}.  Since radio sources grow in size as they get older
(e.g. \cite{scheuer74}, \cite{baldwin82}), GRGs must represent a late
phase in the evolution of radio sources.

Probably not all sources will live long enough, or grow rapidly
enough, to reach the size of the Giant radio sources. What fraction of
sources do and under what circumstances is still unclear. According to
radio source evolution models, GRGs must be extremely old
(i.e. typically above $10^8$ yr) and/or located in very underdense
environments, as compared to smaller radio sources
(e.g. \cite{kaisal99}).  The age of a radio source can be estimated by
sensitive multi-frequency radio observations of the radio lobes
(e.g. \cite{allea87}).  The first systematically obtained results of a
small sample of GRGs show that the spectral ages so found are indeed
comparable to those expected from source evolution models \cite{macket
al98}.  However, such studies have always been severely hampered by
the fact that large uniformly selected samples of GRGs do not exist.

Since GRGs have sizes which are considerably larger than galactic or
even cluster halo cores, their radio lobes interact with the
intergalactic medium (IGM). Therefore, by studying the properties of
the radio lobes we can constrain the properties of the IGM. For
instance, in an adiabatically expanding Universe filled with a hot,
diffuse and uniform IGM, the IGM pressure, $p_{igm}$, should increase
as a function of redshift, $z$, as $p_{igm} \!\propto\! (1+z)^5$
(e.g. \cite{subrsar93}). Using a small sample of GRGs, Subrahmanyan \&
Saripalli \cite{subrsar93} limit the local value of the IGM pressure,
$p_{igm,0}$, to $p_{igm,0} \lesssim 2 \times 10^{-14}$ dyn
cm$^{-2}$. Cotter \cite{cotter98} performs a similar analysis with a
larger sample of sources which also extends to higher redshifts (up to
$z\sim1$), and he confirms that the observed evolution in radio lobe
pressures does not contradict a $(1+z)^5$ relation. However, these
results might be biased since it is likely that the known distant GRGs
are the most powerful ones at that epoch and which thus have the
highest equipartition lobe pressures.

In order to address the above issues more carefully, it is vital to
use a sample of GRGs with well understood selection effects. We have
compiled such a sample of GRGs from the 325-MHz Westerbork Northern
Sky Survey (WENSS; e.g. \cite{rengetal97}). From the WENSS we have
selected all radio sources with a (projected) size exceeding 1 Mpc, a
flux density above 1 Jy, an angular size above 5 arcminutes and a
distance from the galactic plane larger than 12.5 degrees. Our sample
consists of 26 sources, of which 10 are newly discovered GRGs
\cite{schoen99}.

We have used the WENSS radio maps, in combination with maps from the
1.4-GHz NRAO VLA Sky Survey (NVSS, \cite{condetal98}) and our 10.5-GHz
Effelsberg observations to study the properties of the radio lobes of
the 22 FRII-type \cite{fanril74} sources in our sample. Since GRGs are
large sources on the sky, it is already possible to achieve this with
the modest angular resolution of these datasets (i.e. $\sim 1$ arcmin).  In cases where we
could determine the spectral age from a steepening of the lobe
spectrum away from the hotspot, we find ages which are in the range of
50 -- 100 Myr, which agrees with earlier results on GRG spectral ages
(e.g. \cite{macketal98}). We also find that the GRGs tend to have
higher lobe advance velocities than smaller sources of similar
observed radio power \cite{schoen99}.

\section{Radio lobe pressure evolution}

Estimates of the internal pressures of radio lobes are directly
obtained from estimates of the equipartition energy densities, $u_{eq}$,
since in a relativistic plasma the pressure is given by $p =
\frac{1}{3}u$, where $u$ is the energy density. The equipartition
energy density $u_{eq}$ can be obtained from radio observations
(e.g. \cite{miley80}).

If a radio source is well resolved (i.e. larger than 8 arcminute), we
have divided the radio lobe into several regions and estimated the
energy density in each of these.  We have made measured the radio lobe
pressure along the radio axes of the FRII-type sources.  In
Fig. \ref{fig:schoenmakers1} we have plotted four examples of energy
density (i.e. pressure) profiles of GRGs. All sources presented here
show a decrease in energy density when going back from the hotspots
(situated at the outer edges) to the radio cores in the center. In
some cases, the energy density rises again in the vicinity of the
center. This can be either due to the presence of a strong radio core
and/or jet, or to a real increase in the lobe pressure as a result of
a higher pressure in the environment caused by the presence of, e.g.,
a gaseous galactic halo. The decrease in energy density as a function
of hotspot distance indicates the presence of a pressure gradient in
the lobes. This suggests that the radio lobes are still overpressured with respect to the ambient medium, the IGM.

\begin{figure}[t!]
\begin{tabular}{c c}
\includegraphics[width=0.28\textwidth,angle=90]{schoenmakers_11.eps} &
\includegraphics[width=0.28\textwidth,angle=90]{schoenmakers_12.eps} \\
\includegraphics[width=0.28\textwidth,angle=90]{schoenmakers_13.eps} &
\includegraphics[width=0.28\textwidth,angle=90]{schoenmakers_14.eps} \\
\end{tabular}
\caption{\label{fig:schoenmakers1} Energy density profiles along the
radio axes of the GRGs B\,0109+492 (upper left), B\,1312+698 (upper
right), B\,1543+845 (lower left) and B\,2043+749 (lower right).
Numbers along the top denote the distance from the core in kpc.  Note
the increase of the energy density (or, equivalently, pressure)
towards the outer edges of the sources. Also note the increase towards
the center in B\,1543+845 and B\,2043+749; the latter has a strong
radio core causing this effect, but the former has not.}
\end{figure}

\begin{figure}[t!]
\begin{center}
\includegraphics[width=0.5\textwidth,angle=90]{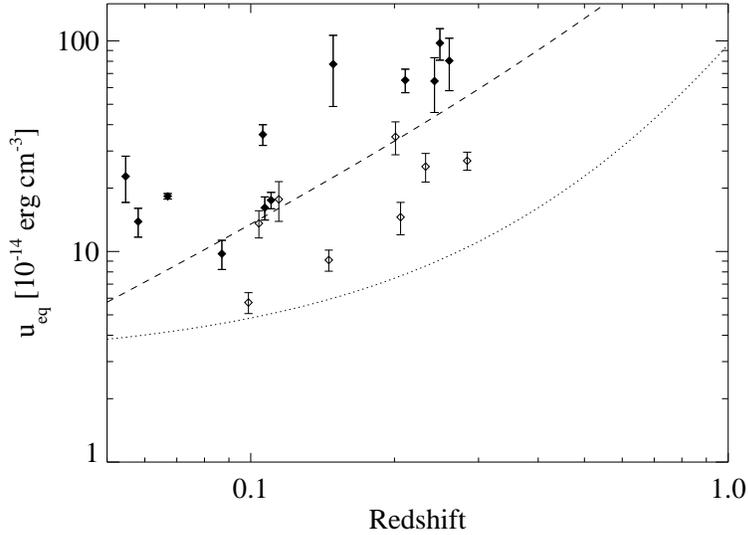} 
\end{center}
\caption{\label{fig:schoenmakers2} A plot of the intensity weighted
average equipartition energy density in the radio lobes of the
FRII-type GRGs in our sample against redshift. Filled symbols denote
sources with a linear size between 1 and 2 Mpc, open symbols are
sources above 2 Mpc. The dashed line indicates the expected energy density
of a source with a flux density, volume and spectral index as given
by the median values of these properties in our sample, and thus gives
the relation which is expected on basis of our selection criteria.
The dotted line indicates the lower limit if the pressure in the lobes
is dominated by relativistic particles, the lobes are in pressure
equilibrium with the IGM and the pressure of the IGM follows the
relation $p_{igm} = 1.0 \times 10^{-14}\cdot(1+z)^5$ dyn cm$^{-2}$.}
\end{figure}

We have also calculated the intensity weighted average lobe energy
density for each source in our sample. Fig. \ref{fig:schoenmakers2}
shows these values plotted against the redshifts of the sources. We
have made a separation between sources which are smaller and larger
than 2 Mpc (closed and open symbols, respectively).  There are two
things to note in this plot. First, the larger sources tend to have
the lowest average lobe pressures. This is a confirmation of the trend
already noted by Cotter \cite{cotter98} for sources smaller and larger
than 1 Mpc. Second, although the redshift range of our sources is
limited ($z \lesssim 0.4$), there appears to be a correlation between
energy density and redshift. This agrees with the results from
Saripalli \& Subrahmanyan \cite{subrsar93} and Cotter \cite{cotter98}
and does not contradict a $(1+z)^5$ increase of the IGM pressure,
provided that the current day value is $\lesssim 10^{-14}$ dyn
cm$^{-2}$ (indicated by the dotted line in
Fig. \ref{fig:schoenmakers2}).  

We note, however, that the observed increase in lobe energy density
with increasing redshift in Fig. \ref{fig:schoenmakers2} also exactly
matches the expected behaviour for a source of fixed dimensions and
flux density. This is shown by the dashed line in Fig.
\ref{fig:schoenmakers2}, which indicates the expected equipartition
energy density in the lobes of a source with a size equal to the
median size of the GRG sample and a flux density equal to the median
flux density of the sample. The slope of this line matches the
observed redshift relation of the energy density in our sources.
Therefore, the observed redshift relation is more likely due to the
use of a flux density and source volume limited sample, than to any
cosmological effect. The same relation must apply to the sources of
Cotter \cite{cotter98}, since his sample of high redshift GRGs also
form a flux density limited sample (flux density between 0.4 and 1.0
Jy at 151 MHz) of sources larger than $\sim 1$ Mpc. We conclude that
there is therefore no evidence for a strong increase in the IGM
pressure with increasing redshift, although we cannot reject it
either.

To investigate whether the IGM pressure truly evolves as strongly as
$(1+z)^5$, it would be necessary to find Mpc-sized radio sources at
high redshifts.  From Fig. \ref{fig:schoenmakers2} it can be deduced
that the existence of a population of sources with lobe energy
densities of $\lesssim 3 \times 10^{-14}$ erg cm$^{-3}$ at redshifts
of at least 0.6 would be difficult to reconcile with a strong pressure
evolution, unless the current day IGM pressure is much lower than
$10^{-14}$ dyn cm$^{-2}$. Such sources are expected to have flux
densities of $\lesssim 200$ mJy at 325 MHz (assuming a size of 1.5
Mpc, a spectral index of $-0.8$ and a aspect ratio of 3), and are thus
detectable by WENSS. Indeed, we have started to compile a sample of
such sources. Detailed observations of their radio structures at low
frequencies (i.e. $\sim 100$ MHz) will be necessary to correctly
estimate the lobe energy densities.

\section{'Double-double' radio galaxies}

One of the outstanding issues concerning extra galactic radio sources
and other Active Galactic Nuclei (AGN) is the total duration of their
active phase. For radio sources, this physical age of the nuclear
activity is not to be confused with the radiative loss age determined
from radio spectral ageing arguments; many extra galactic radio
sources probably have a physical age well surpassing their radiative
loss age (e.g. \cite{laanper69}, \cite{eilek96}, but also see
\cite{parmaetal99}).  The length of the active phase is intimately
related to the possible existence of duty cycles of nuclear
activity. In case nuclear activity is not continuous, how often do
interruptions occur and how long do they last?

Such duty cycles can only be recognized if there is some mechanism to
preserve the information of past nuclear activity for a long enough
time to be recognized when a new cycle starts up.  In extended radio
sources, such a mechanism is potentially provided for by the radio
lobes, since they remain detectable for a long time after their energy
supply has ceased (possibly up to a few $10^7$ yr;
e.g. \cite{komgub94}).  If a new phase of activity should start before
the `old' radio lobes have faded, and if this activity manifests
itself by the production of jets, we can in principle recognize this
by the observation of a new, young radio source embedded in an old,
relic structure. One well-known candidate for such a `restarted' radio
source is the radio galaxy 3C\,219 (\cite{clarkbur91},
\cite{clarketal92}, \cite{peretal94}). In this source, radio jets have
been observed which abruptly become undetectable at some point between
the core and the leading edge of the outer radio lobes. However,
sources such as this are extremely rare and difficult to recognize.

\begin{figure}[t!]
\begin{center}
\includegraphics[width=\textwidth]{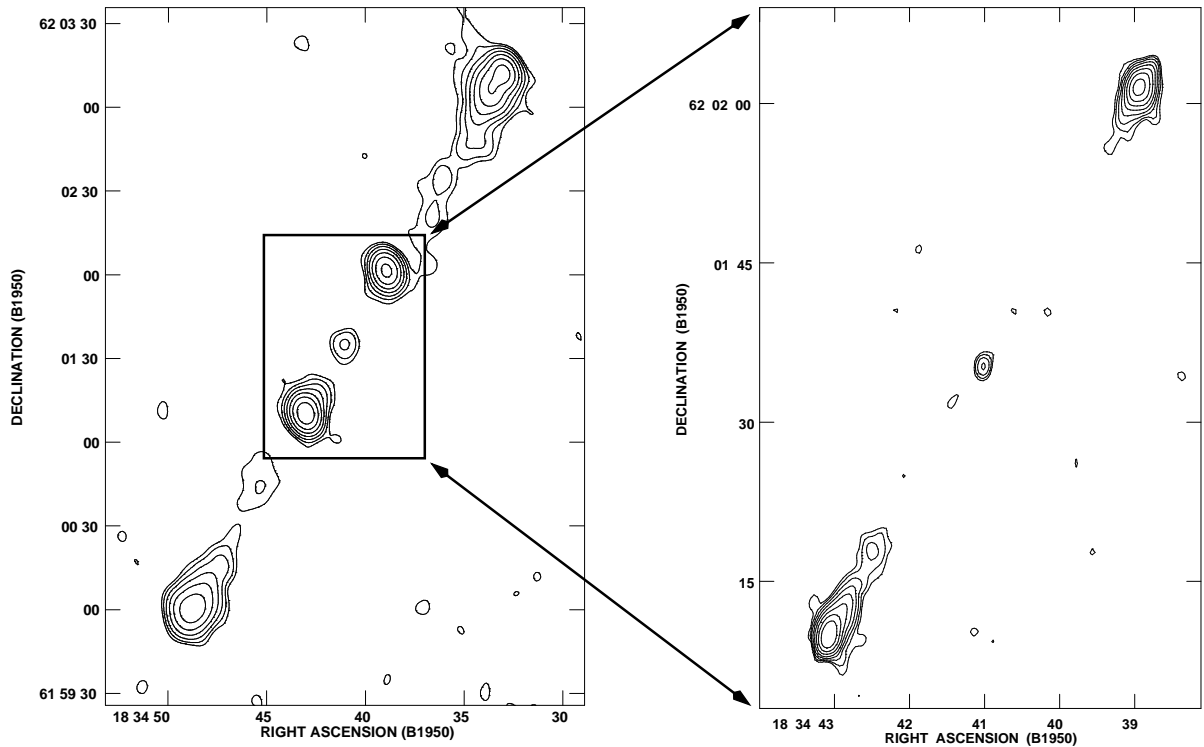} 
\end{center}
\caption{\label{fig:schoenmakers3} A radio contour plots of the DDRG
B\,1834+620. On the left we present a map from 8.4-GHz VLA
observations (8 arcsec resolution), which clearly shows the outer lobe
structures. The map on the left shows the inner structure at a higher
resolution of $\sim 1$ arcsec and at a frequency of 1.4 GHz. Clearly,
the morphology of the inner structure is that of an FRII-type radio
source. The overall size of the outer structure is 1660 kpc, that of
the inner structure 420 kpc. The redshift of this source is
0.519\,. Contours are drawn at intervals of a factor 2 starting at
0.22 mJy beam$^{-1}$ (8.4 GHz) and 0.17 mJy beam$^{-1}$ (1.4
GHz). Dashed contours denote negative levels.}
\end{figure}

\begin{figure}[t!]
\begin{center}
\includegraphics[width=\textwidth]{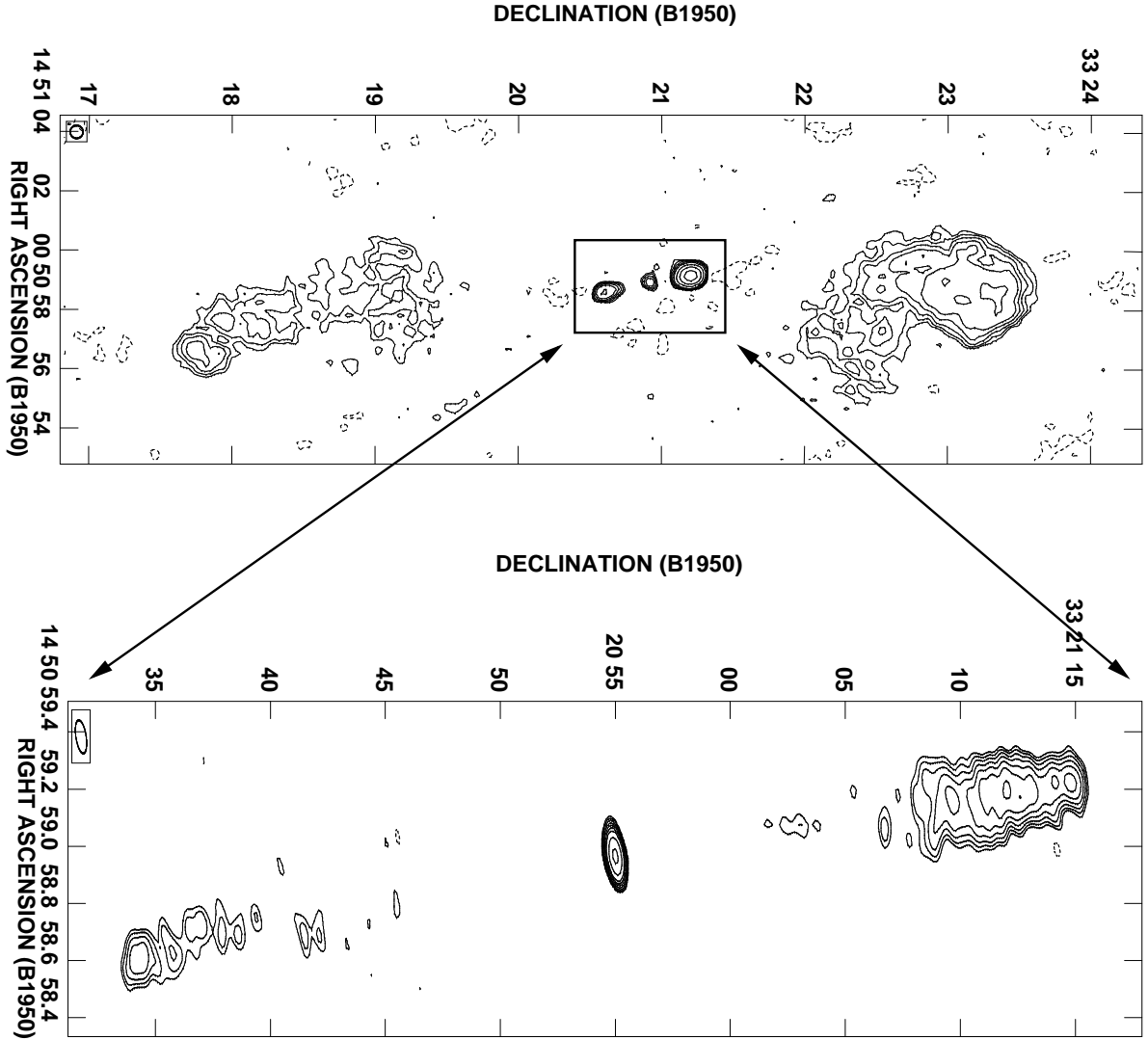} 
\end{center}
\caption{\label{fig:schoenmakers4} A radio contour plots of the DDRG
B\,1450+333. At the top we present a map from the 1.4-GHz VLA FIRST
survey \cite{becketal95}, at the bottom we shows the inner
structure at a higher resolution and at 4.9 GHz. Clearly, the
morphology of the inner structure is that of a FRII-type radio
source. The overall size of the outer structure is 1680 kpc, that of
the inner structure 180 kpc. The redshift of this source is
0.249\,. Contours are drawn at intervals of a factor $\sqrt{2}$
starting at 0.45 mJy beam$^{-1}$ (1.4 GHz) and 0.12 mJy beam$^{-1}$
(4.9 GHz). Dashed contours denote negative levels.}
\end{figure}

During our search for GRGs in the WENSS survey, we have found several
sources which are excellent candidates for restarted radio
sources. Radio contour plots of two of these, B\,1834+620 and
B\,1450+333, are shown in Figs. \ref{fig:schoenmakers3} and
\ref{fig:schoenmakers4}, respectively. Both cases are clearly
different from `standard' FRII-type radio galaxies. Since they consist
of an inner double-lobed radio structure as well as a larger outer
double-lobed structure, we have called these sources `double-double'
radio galaxies (DDRGs;\cite{schoen99}, \cite{schoenetal99a},
\cite{schoenetal99b}).

In Schoenmakers et al. (\cite{schoen99}, \cite{schoenetal99a},
\cite{schoenetal99b}) we present a small sample of seven of these
peculiar sources. Among the general properties are that in all cases
the inner structures are less luminous than the outer structures, and
that the difference in radio power between the inner and outer
structures appears to decrease with increasing size of the inner
structure. Further, almost all sources in our small sample have large
linear sizes, above 700 kpc and ranging up to 3 Mpc.

The observed two-sidedness and symmetry in the morphology of the inner
structures strongly suggests a central cause for this phenomenon, and
we believe that an interruption of the central jet-forming activity is
the most likely one. For the source B\,1834+620
(Fig. \ref{fig:schoenmakers3}) we are able to constrain the time-scale
of the interruption to $\lesssim 6$ Myr (\cite{schoenetal99b},
\cite{schoenetal99c}). What actually causes the AGN to interrupt the
radio activity is unclear. Possible options are a large inflow of gas
into the central region of the galaxy (e.g. by an infalling large
molecular cloud), causing an instability in the accretion flow onto
the central massive black hole.

One of the most fascinating aspects of these sources is the actual
existence of the inner radio lobe structure. Numerical simulations of
restarting systems (e.g. \cite{clarketal92}) and physical
considerations on the properties of cocoons produced by jets agree in
that the density inside cocoons are not high enough to allow the
formation of strong shocks, related to the formation of hotspots and
radio lobes. The fact that we nevertheless observe these must indicate
that the density inside the cocoons is much higher than predicted by
these models. Kaiser et al. (\cite{kaiseretal99}; also see
\cite{schoen99}) present a model in which the density inside the
cocoon is actually increased as the result of the entrainment and
subsequent shredding of warm clouds in the IGM by the expanding
cocoon. They show that after a long enough time (i.e. a few $10^7$ yr)
the density inside the cocoon may have increased sufficiently to allow
a new system of lobes and hotspots to be formed after an interruption
of the jet flow. The long time scale can explain the large size of the
DDRGs. The low densities inside the cocoon as compared to the ambient
medium can explain the low radio power and the high advance velocities
of the inner structures (estimated to be 0.2$c$ -- 0.3$c$
(\cite{schoen99}, \cite{schoenetal99b}, \cite{schoenetal99c}).

We therefore argue that the DDRGs show a distinct phase in the
evolution of radio sources. Among the questions that remain are the
following: How many radio sources actually go through such a phase?
What is the cause of the interruption? To answer these questions much
more detailed studies of these fascinating sources are required. To
investigate the rate of occurrence, it might be interesting to search
for old relic structures around known radio sources. Such an
undertaking must be performed at low frequency, with high sensitivity
and dynamic range. With the proposed sensitivity of SKA, this should
be a feasible project.  The cause of the interruption can perhaps be
investigated by detailed optical and kinematical studies. However, the
chance of finding anything may actually be small if the cause is only
due to small-scale events such as an infalling cloud.
 
\section{The importance of SKA}

The next important step in GRG research will be the compilation of a large
sample of higher redshift GRGs. This is interesting in many respects:
First of all, such a sample will provide us with important constraints
on the cosmological changes in radio source evolution. Since this is
closely related to the evolution of the environments of radio sources
on scales up to a few Mpc, this can teach us more about the coupling
between the evolution of galaxies in clusters and the intra-cluster
gas. Second, sensitive high resolution observations are needed to
investigate the radio lobe properties of high redshift GRGs in some
detail. This is vital to obtain information on the spectral ages of
GRGs, and the ageing processes themselves. Since the energy density of
the microwave background increases as $(1+z)^4$, the effect of Inverse
Compton scattering on the ageing of the particles in the radio lobes
becomes increasingly more important toward higher redshift. This will
make the radio spectra of the bridges in the lobes steepen
considerably, allowing only sensitive low frequency observations to
detect these faint regions. Detailed studies of the rotation measures
towards the radio lobes can show us density structures in the ambient
medium of the lobes at distances of a Mpc from the host galaxy. Also, if
part of the Faraday rotation were to occur within the radio lobes, such
a study may yield unique information on the internal properties
of the radio lobes, such as the thermal particle density and the magnetic
field strengths.

In case of the DDRGs similar studies can reveal the properties of the
medium around the inner lobe structures, and can thus play an
important role in testing the model proposed by Kaiser et
al. \cite{kaiseretal99} for the formation of these structures.
Another important topic is how common the DDRG phenomenon is, a
question which is closely related to that of the occurrence of
duty-cycles in AGN. The DDRGs which we know now are the ones with the
most prominent outer structures and as such that may only form the tip
of the iceberg of radio sources with multiple periods of
activity. With a sensitive low frequency telescope we can investigate
the areas surrounding known radio sources for possible relic
structures, indicative of an earlier phase of activity.
 
Therefore, the properties that will make SKA an extremely important
instrument for future research of GRGs and DDRGs are a high
(sub-arcsecond) angular resolution at a low frequency (100 -- 1000
MHz), combined with excellent sensitivity and polarization
characteristics. In order to be able to investigate such large
structures as GRGs are, SKA should be capable of mapping large
structures on the sky, up to a few tens of arcminute, without losing
sensitivity. We therefore argue for a next generation radio telescope
(i.e. SKA) which consists of a combination of a central compact array
(few km diameter) and several long baselines (preferably up to a
thousand km), which should be able to observe routinely at frequencies
as low as 50--100 MHz.

\section*{Acknowledgements}
KHM is supported by the Deutsche For\-schungs\-gemein\-schaft, grant
KL533/4--2 and by the European Commission, TMR Programme, Research
Network Contract ERB\-FMRX\-CT96-0034 ``CERES''. This work is supported in
part by the Formation \& Evolution of Galaxies network set up by the
European Commission under contract ERB-FMRX-CT96-086 of its TMR
programme.

\end{document}